\pdfoutput=1

\documentclass[sigconf]{acmart}

\AtBeginDocument{%
  \providecommand\BibTeX{{%
    \normalfont B\kern-0.5em{\scshape i\kern-0.25em b}\kern-0.8em\TeX}}}

\copyrightyear{}
\acmYear{}
\acmDOI{}

\acmConference[]{}
\acmBooktitle{}
\acmPrice{}
\acmISBN{}

\usepackage{multirow}
\usepackage{multicol}
\begin{document}

\title{Detecting Radical Text over Online Media using Deep Learning}

\author{Armaan Kaur}
\email{armaankaurgill@gmail.com}
\affiliation{%
  \institution{Cyber Security Research Centre, Punjab Engineering College, (Deemed To be University )}
  \streetaddress{Sector 12}
  \city{Chandigarh}
  \country{India}
  \postcode{}
}

\author{Jaspal Kaur Saini}
\email{sainijassi87@gmail.com}
\affiliation{%
  \institution{Cyber Security Research Centre,Punjab Engineering College, (Deemed To be University )}
  \streetaddress{Sector 12}
  \city{Chandigarh}
  \country{India}}

\author{Divya Bansal}
\email{divya@pec.edu.in}
\affiliation{%
  \institution{Cyber Security Research Centre,Punjab Engineering College, (Deemed To be University )}
  \streetaddress{Sector 12}
  \city{Chandigarh}
  \country{India}}

\renewcommand{\shortauthors}{Armaan et al.}

\begin{abstract}
  Social Media has influenced the way people socially connect, interact and opinionize. The growth in technology has enhanced communication and dissemination of information. Unfortunately, many terror groups like jihadist communities have started consolidating a virtual community online for various purposes such as recruitment, online donations, targeting youth online and spread of extremist ideologies. Everyday a large number of articles, tweets, posts, posters, blogs, comments, views and news are posted online without a check which in turn imposes a threat to the security of any nation. However, different agencies are working on getting down this radical content from various online social media platforms. The aim of our paper is to utilise deep learning algorithm in detection of radicalization contrary to the existing works based on machine learning algorithms. An LSTM based feed forward neural network is employed to detect radical content. We collected total 61601 records from various online sources constituting news, articles and blogs. These records are annotated by domain experts into three categories: Radical(R), Non-Radical (NR) and Irrelevant (I) which are further applied to LSTM based network to classify radical content. A precision of 85.9\% has been achieved with the proposed approach.
\end{abstract}



\maketitle
\section{Introduction}

Different social media platforms follow different approach of connecting people and sharing information. With time, social media has penetrated and integrated into our lives so well that we share all the personal data and check-ins online.  Such is the extent that it has opportunely influenced the opinions and decision-making of consumers. Reachability to masses has been expedited by the advancement in Internet facilities.  It is startling how people leave word about their daily activities using social media and share information that actually affects the users connected. Hence, users with similar intents come aboard
 online\cite{java2007we}. 
 \par A major drawback of such advancements is spreading of radicalization has eventually migrated from physical interactions at public places to virtual conversations over social media platforms.  At present, the Islamic State (IS) is indisputably the leading organization in spreading of their propaganda online\cite{fernandez2018understanding}. The extremist groups spread their propaganda online, influence vulnerable frustrated individuals, motivate towards their ideology, recruit online, raise funds from supporters, instigate people to become members of aggressive communities and provoke lone-wolf terrorist into acts of violence. Based on the studies reviewed by Hassan et al. \cite{hassan2018exposure}, contingent evidences have been provided in favour of fact that exposure of a person to radical content poses higher risks of engaging in political violence. Active seekers are at greater threat than passive ones. \par Doosje et al. defines radicalization as:
    "Radicalization is a process through which people become increasingly motivated to use
violent means against members of an out-group or symbolic targets to achieve behavioral
change and political goals.\cite{doosje2016terrorism}"
The model of radicaization is summarized in \cite{doosje2016terrorism} undergoing three phases: (1) Sensitive phase: Being sensitive to radical ideas. (2)
Membership phase: Becoming part of a radical group. (3)
Action phase: Act on behalf of the ideologies of the radical group. 
Apart from recruiting attackers, intermediaries are also recruited online for carrying messages and spreading martyr videos. The major threats due to wide spread of extremist ideas over the internet are enlisted below:  
\begin{itemize}
\item Recruitment of ISIS fighters and searching for lone wolf terrorists online.
\item Rapid dissemination of extremist ideologies.
\item Identification of outraged, unemployed, anti-nationalist youth online that can be easily influenced.
\item Many other jihadi groups have started adopting ISIS's method of mobilizing supporters globally through social media.
\item The establishment of online jihadi community may accelerate more rapidly with increasing global radicalization.
\item Greater penetration of terror groups into social media accounts to target youth.
\item Online donations and fund raising activities are becoming more digital and ease to reach masses over social media makes such activities spread easily..
\end{itemize}
The increasing role of Internet in contemporary terrorism has persuaded the world leaders and governments to call on social media companies to put more efforts in getting rid of violent and extremist content floating on their platforms. This would further hamper freedom of speech, therefore targeted deletion requires an effective analysis of content posted online.
 With the upsurge in development of radicalization, it is important to devise ways for detection and keep a technical edge with advancement in technologies. Also, Mukherjee et al.\cite{mukherjee2012mining} refers to the limited research in the area of exploring texts of discussion forums deeply. As many such platform are used for contentious issues such as political, social and religious, hence these are topics of discussion and often go unaddressed by governments but it is critical to recognise them. Topic modelling using maximum entropy is one of the text analytics technique utilised in previous work \cite{mukherjee2012mining} and compared with LDA and SLDA.  
\par Our work proposes a novel approach in detection of radicalization on online media that put to use Deep Neural Networks using word embeddings. Exploring the field of Deep Learning is worthwhile that can certainly automate the process of extraction of complex data representations. These algorithms have a layered and hierarchical structure for learning and represention of data,
where low level attributes are used to define high level attributes. \cite{deeplearning}. Mudgal et al. apply entity matching with deep learning to look for the prospects of this much less explored area and finds out that Deep Learning has convincing results.It can easily outperform in the absence of data cleaning and therefore, push boundaries of current automated tasks\cite{mudgal2018deep}. The emerging field of deep learning calls for extensive research \cite{clstm}. 

\par This paper presents an automated approach to detect radicalisation over online social media. We utilised text analytics and deep learning mechanism to identify radical content. We started with studying various research evidences which throws light on application domains of deep learning techniques in text analytics. A comparative study of work done on multiple datasets is presented in current work. LSTM based feed forward neural network is employed to detect radical content. Further sections discuss the key details of our work.    
\section{Related Work}
For the purpose of literature survey, different papers were studied to have an insight into development of radicalization, techniques used to detect it and varied deep learning algorithms applied onto text analytics.
\begin{table}[h!]
\small
\centering
 \begin{tabular}{|p{1.25cm} |p{1.75cm} |p{1cm}| p{2.85cm}|} 
 \hline
 \textbf{Study} & \textbf{Technique} & \textbf{Source of data} & \textbf{Precision} \\ [0.5ex] 
 \hline\hline
 Fernandez et al.(2018) & Context modelling based on ontologies and knowledge-bases, NB, SVM, J48 & Twitter & 0.859 \\ [0.5ex]
\hline
Barhamgi et al.(2018) & Ontology-based semantic description & Twitter & 0.77(Ontology), 0.38(Baseline) \\ [0.5ex]
\hline
 Njagi Dennis Gitari et al.(2015) & lexicon-based approach & web forums, blogs, comment section of news reviews & Subjective sentence:
 
 67.21(sem), 71.22(sem+hate), 73.42(sem+hate+theme) Without subjective sentence: 
 
 58.42(sem), 63.24(sem+hate), 65.32(sem+hate+theme) \\[0.5ex]
 \hline
 Swati Agarwal et al.(2015) & SVM, KNN & Twitter & 0.48(KNN), 0.78(SVM) \\ [0.5ex]
\hline
 M. Munezero et al.(2013) & Bag-of-words, Ontology-based emotion description, SVM, Multinomial NB, J48(DEcision tree) & Movie review, ISEAR, wikipedia article & SVM: 
 
 0.891(ASB+ISEAR), 0.955(ASB+MovRev), 0.803(ASB+Wiki), 
 
 J48: 
 
 0.832(ASB+ISEAR), 0.966(ASB+MovRev), 0.821(ASB+Wiki), 
 
 MNB: 
 
 0.896(ASB+ISEAR), 0.996(ASB+MovRev), 0.908(ASB+Wiki) \\ [0.5ex]
 \hline
 Pir Abdul Rasool Qureshi et al.(2011) & DETECT framework & blogs, discussion forums, news articles, etc & NA \\[1ex] 
 \hline
\end{tabular}
\linebreak
\caption{Comparison of studies based on their techniques, sources of data and accuracies.}
\label{tableradical}
\end{table}
\par
The comparison in Table \ref{tableradical} provides overview of techniques used for data radicalization along with the precision score achieved. As can be implied from Table \ref{tableradical}, twitter can be used as a  source for text analytics if annotated well and has been widely used in most of the studies.
 Fernandez et al.\cite{fernandez2018contextual} analysed 114k tweets (17k from pro-ISIS users and 97k from general ones) and achieved an improvement of 4.5\% with context modelling over keyword-based search. Mahmoud Barhamgi et al.\cite{barhamgi2018social} collected twitter messages exchanged between 20 accounts tagged radical by Kaggle dataset that had total 2317 messages both radical and neutral posted by selected users. Swati Agarwal et al. \cite{agarwal2015using} combined datasets: UDI TwitterCrawl (Aug2012) and ATM TwitterCrawl (Aug2013), followed by filtration using language detection library to be left with 45.3 million english tweets. The tweets were manually annotated as hate and extremism promoting tweets and semi-supervised learning followed as the learning was performed by training the model over data of one class only. Overall LibSVM can be found to outperform KNN.
 \par
The classification model can also be trained over dataset collected from multiple sources like blogs, discussion forums, news and articles as all these sources have opinions of users. Apache Nutch has been extended into framework proposed in \cite{qureshi2011detecting} for acquisition of open source information like blogs, discussion forums, news articles, facebook, data provided by analysts, etc. Njagi Dennis Gitari et al. \cite{gitari2015lexicon} has chosen web forums, blogs and comment sections of news reviews as source of data. The corpus is divided into two: first one consisted of 30 blogs and second one consisted of 150 pages  having paragraphs from documents, labelled as strongly, weakly or not hateful. The classification is based on a lexicon built using semantic, hate and theme-based features from dictionary and corpus both. The precision improved with addition of hate and thematic elements to semantics and also further with the use of subjective sentences. The results for strongly hateful sentences are shown in the table.\par
M. Munezero et al. \cite{munezero2014automatic} collected corpora from antisocial behaviour texts (ASB), Movie Reviews and Wikipedia extracts. ISEAR corpus used for emotion description. WEKA tool is used for the implementation of machine learning algorithms: SVM, MNB, and J48 decision trees. In most of the cases, the precision can be seen to have improved with addition of emotions. The MNB has been found to have performed better compared with SVM and J48 when bag-of-words features combined with emotions. The addition of sentiment analysis to extend the work has been suggested.\par
Table \ref{tableradical} presents various techniques used for detecting radicalization. It is seen from the table that deep learning algorithms has not been utilised in the field so far.
As can be seen, by and large the performance of classification algorithms depends upon goodness of data representation. The task of feature engineering consumes a lot of effort and directly impacts the performance of machine learning algorithm. Deep learning is one such area of research that dwells on automated extraction of features and representation of data, thus capturing complex patterns observed \cite{deeplearning}. Further a comparative analysis of the deep learning mechanisms used in various applications in text analytics domain studied is shown in Table \ref{tablednn}. 
  
\begin{table}[!h]
\small
    
    \begin{tabular}{|p{1.75cm}|p{3cm}|p{3cm}|} 
     \hline
     \textbf{Study} & \textbf{Technique} & \textbf{Dataset} \\  [0.5ex] 
     \hline\hline
     Zhou et al.(2015)  & C-LSTM Neural Network & Stanford Sentiment Treebank (SST) consisting of movie reviews \\ [0.5ex]
     \hline
     Zhou et al.(2016)

 & BLSTM-2DPooling on 6 different classification tasks & Stanford Sentiment Treebank dataset \\ [0.5ex]
     \hline
     Xu et al.(2017)

 & Naive Bayes, LSTM and Social Network Analysis & Twitter data of a terrorist organization and supporting communities \\ [0.5ex]
     \hline
     Rybinski et al.(2018) &
 Logistics regression using bag-of-words and n-gram schemes compared with DNN operating on word embeddings  & Toxic Comment Classification
Challenge dataset \\ [0.5ex]

 \hline
Tommasel et al.(2018)

 & SVM and RNN to analyze different feature sets & Dataset presented in Kumar et al.\cite{kumar2018aggression} 
     \\[1ex] 
          \hline

    \end{tabular}
    \caption{Comparison of studies on text analytics using deep learning on the basis of technique and dataset used.}
    \label{tablednn}
\end{table}
\par
Zhou et al.\cite{clstm} aims to uphold word orders by extracting n-grams features using a CNN layer over an LSTM layer that yields sentence representations over a Stanford Sentiment Treebank dataset of 11855 movie reviews. The best accuracy is achieved with C-LSTM in fine grained classification (49.2\%), with Bi-LSTM in binary classification (87.9\%) and with C-LSTM in question classification (94.6\%). Besides Zhou et al. in his another work \cite{blstm} proposes BLSTM-2DPooling mechanism which is also applied over Stanford Sentiment Treebank dataset but conducts six cases of text classification, namely subjectivity classification, newsgroups classification, question classification, and sentiment analysis. Here, the text is transformed into vectors using a Bidirectional LSTM and successfully capturing information from past and future, followed by 2D max pooling  utilized in obtaining 
fixed-length vectors. Finally, 2D convolution  captures meaningful features to be represented from input text. The best accuracy achieved by author is 89.5\%. \par
The artificial intelligence techniques can also be combined with Social Network Analysis task for an exhaustive study \cite{aisna} where a terrorist organization is targeted. The 33,000 tweets were used as positive corpus obtained from this organization and 20,000 tweets were used as negative corpus procured from an NLTK toolkit. Naive Bayes classifier used word frequencies as features assuming texts are independent of each other compared to LSTM that considers them interrelated and used word embeddings. With the increase in training set proportion, LSTM accuracy surpassed Naive Bayes remaining around 96.5\%  and performed better by exceeding 99\%.
\par
Rybinski et al. \cite{toxicity} have compared traditional methods like logistics regression using bag-of-words and n-gram schemes with DNN operating on word embeddings in the task of predicting the toxicity class to which input text belongs. It is conducted over Toxic Comment Classification Challenge dataset and attained great results with neural networks achieving a mean ROC-AUC score that transcends value of 0.99. Tommasel et al. \cite{aggression} have used SVM and RNN to analyse 4 different feature sets: Word embeddings using GloVe, sentiment features using SentiWordNet,  TF-IDF with punctuation features and TF-IDF coupled with n-grams. The dataset used is presented in Kumar et al. \cite{kumar2018aggression} that has posts related to hashtags and pages used over Twitter and Facebook by Indians. The best  accuracy of 81\% is achieved when  features set with  count of curse words and their intensities is used.
\begin{figure*}[!h]

	\includegraphics[width=\linewidth]{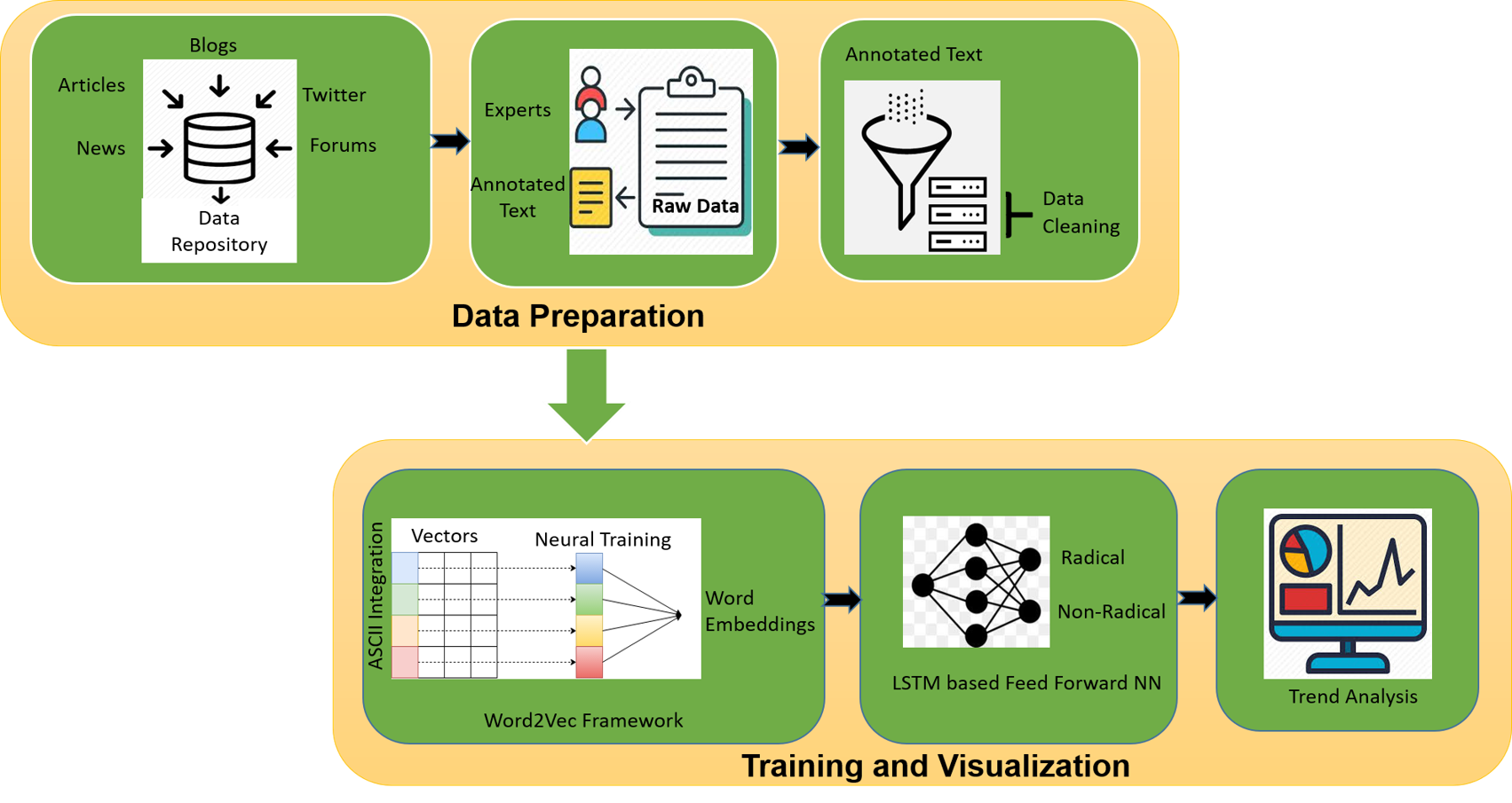}
	\caption{Detailed Architecture of the Proposed Approach}
	\label{fig1:architceture}
\end{figure*}
 
\section{Proposed Work}
The proposed architecture is as shown in Figure \ref{fig1:architceture}, divided into two parts: (1) Data preparation comprising of data collection, annotation and cleaning the raw data. (2) Training and Visualisation comprising of generating word embeddings, then classifying the texts as radical or non radical, followed by trend analysis. The details of each component is described in detail in the following sections: 

\subsection{Data Preparation}
\begin{enumerate}
	\item \textbf{Data Collection}
	 \\The data is collected from multiple sources namely: Blogs, Articles, Twitter, News. We have taken expert guidance to identify data sources that may contain radical content. Details of the same is shown in Table \ref{tabledata} below.
	 \begin{table} [ht!]
\small
    \begin{tabular}{|l|l|l|l|} 
     \hline
     \textbf{Type} & \textbf{Name} & \textbf{Time Period} & \textbf{Records} \\ [1ex] 
     \hline\hline
     \multirow{4}{5em}{News} & Greater Kashmir & 2018 & 52 \\
& Kashmir Reader & 2014-2018 & 35637 \\ 
& Kashmir Monitor & 2018 & 1403 \\
& Tribune & 2016-2018 & 20994 \\[1ex]
    \hline
    \multirow{2}{5em}{Articles} & Huriyat & 2010-2013 & 62 \\
    & Huriyat & 2017-2019 & 591 \\ [1ex]
    \hline
     \multirow{7}{5em}{Blogs} & JKLF World & 2009-2018 & 478 \\
     & JKPFL & 2010-2015 & 84 \\
     & Kashmir Truth be Told & 2006-2010 & 145 \\
     & Peoples League & 2009-2016 & 276 \\
     & Sameer Bhatt & 2006-2017 & 475 \\
     & Tanveer and Kashmir & 2008-2019 & 1343 \\
     & United Kashmir & 2008-2014 & 61 \\[1ex]
      \hline
   \multicolumn{3}{|l|}{Total Records} & 61601 \\ [1ex]
    \hline
    \end{tabular}
    \linebreak
    \caption{Details of data collected.}
    \label{tabledata}
    \end{table}
  We developed crawler that collected 61601 records from online social media sources. 
  \begin{table}
\small

 \begin{tabular}{|p{7cm}| p{0.75cm}|} 
 \hline
 \textbf{Sample Text} & \textbf{Label} \\ [0.5ex] 
 \hline\hline
 Srinagar: All Parties Hurriyat Conference, Chairman Syed Ali Geelani On Saturday Said That Forced And Military Occupation Of India Is The Root Cause Of All Miseries And Problems And Added That Authorities In New Delhi Always Applied Its Shrived Tactics To Supress Our Sentiments.
 & R \\ 
 \hline
MUZAFFARABAD: United Jehad Council Has Said That It Has Full Faith And Confidence On Joint Resistance Leadership And Venerates Them. In A Statement To Media, Syed Sadakat Hussain, The Spokesperson Of United Jehad Council Said A Confusion And Misunderstanding Were Being Advertently Or Inadvertently Created After UJC Under A Well Chalked Out Strategy Issued A Schedule From July 8 To 13. "Now UJC Has Decided To Restrict This Schedule To Base Camp Only While On Ground In Kashmir, Joint Resistance Leadership Will Take A Final Call," The Spokesperson Said.
 & NR \\
 \hline
 No more comment moderation,  Dear friends,,  I will be away for a few months because of some issues that need my attention. I have therefore removed the requirement for comment moderation. All posts will appear instantaneously. The only difference is that "anonymous" posts are not possible now. Any registered user (Gmail, Google, OpenID, facebook, wordpress, blogger etc) will be able to post instantaneously. 
 & I 
 
 \\ [1ex] 
 \hline
\end{tabular}
\linebreak
\caption{Sample annotations without preprocessing}
\label{tableannotate}
\end{table}
	\item \textbf{Data Annotation}
\\	For the purpose of data pre-processing and annotations, we defined radicalization as a general term for the online activities and discussions over online social media which support terrorism in the Indian subcontinent by the name of Jihad/Islam. It can be seen as course of action that provokes vulnerable users towards violent extremism and propagating anti-nationalism which is against the interests of the nation. The various contours of radicalization can be summarized as:
\begin{itemize}
    \item Abusing the Indian Army Troops, politicians and using insects as metaphors for them.
    \item Anti-national speeches with trigger factors
    \item Justification of terrorism by the name of Islam or Jihadism or Mujahideen
    \item Terrorists who died while fighting or got assassinated being termed as martyrs and said to have sacrificed their lives
    \item Motivating readers to support extremism 
\end{itemize}
 Some sample annotations are enlisted in the Table \ref{tableannotate}. We provided collected records to two domain experts with definitions of radicalisation and requested them to annotate the given records with following instructions:
\begin{enumerate}
    \item You are provided data from news, articles, blogs and discussions.
    \item Read it carefully and label the data into three categories: 1. Radical (R) 2. Non-Radical (NR) 3. Irrelevant (I)
    \item Label the data as Radical (R): Based on the presence of above-mentioned features of radicalisation devised for the purpose of annotation, label the data as Radical (R). 
    \item Label the data as Non-Radical (NR): All posts which are relevant to the topic of radicalisation but fails to meet the requirements of our definition mentioned above i.e. the post which does not provoke or motivate or justify the topic of radicalisation will be labelled as (NR). For example: related facts about any such radical event like reporting of any event, assassination, arrest or violent killings etc that talks about nation, government, politics, terrorism and conflicted areas in Indian interests.  
    \item Label the data as Irrelevant (I): Label the data as neutral that is completely off the topic or including advertisement.
\end{enumerate}

\subsection{Cohen's Kappa Coefficient}
Cohen's Kappa is a measure of inter-expert agreement between two experts/raters on our annotated data. Cohen's kappa assesses the agreement between two raters classifying items(n) into mutually exclusive classes(c). Cohen's kappa coefficient $(\kappa)$ determines inter-rater agreement for qualitative items that can be categorized. It is calculated as:
\begin{equation}
    \kappa = \frac{p_o - p_e}{1 - p_e} = 1 - \frac{1 - p_o}{1 - p_e}
\end{equation}
where,
\\$p_o$ can be termed as relative observed agreement among experts (similar to accuracy),
\\$p_e$ can be termed as hypothetical probability of chance agreement,
\\using the data observed for calculating the probabilities of each observer by randomly seeing each category.
\begin{table}[!h]
\small
    
    \begin{tabular}{|p{3cm}|p{3cm}|} 
     \hline
     \multicolumn{2}{|c|}{\textbf{Cohen's Kappa Calculation}} \\  [0.5ex] 
     \hline\hline
     Subjects & 1274 \\ [0.5ex]
     \hline
     Experts & 2 \\ [0.5ex]
     \hline
     Kappa coefficient & 0.796
     \\[1ex] 
          \hline

    \end{tabular}
    \caption{Kappa Coefficient values.}
    \label{tablekappa}
\end{table}
\par The value for $\kappa$ comes out to be 0.796 which is a sign of substantial agreement among two raters as shown in Table \ref{tablekappa}.  Considering the significant value of kappa coefficient, we further moved on to utilising this annotated data to detect radicalisation using our proposed approach.   
	\item \textbf{Data Preprocessing}
	\\
After collecting data, we further pre-processed text in order to train our model. We selected records only from articles and blogs as they largely comprised of radical text. In all 1274 records are used for training and validation labelled into 3 categories: R(radical), NR(non-radical), I(Irrelevant). The following steps are taken for cleaning texts: 
\begin{enumerate}
\item All of the documents are in English text apart from the use of some non-English words like 'Azadi' which have been left intact assuming they are readable by the English language speakers.
    \item The records with blank texts or with only punctuations are eliminated while annotation.
    \item The ommission of all the irrelavant punctuations due to absence of any major contribution.
    \item The stopwords removal based upon the list of stopwords maintained depending upon the requirement so as to eliminate the overhead of storing vectors and weight calculation of all the texts in the corpus.
\end{enumerate}
\end{enumerate}
\subsection{Training and Visualization}
\begin{enumerate}
	\item \textbf{Generating Word Embeddings}
	\\All the word embedding representation methods are based on a hypothesis that words generally carry similar meaning that occur in similar contexts. \cite{harris1954distributional} A good representation of text should aim at maintaining the analogy between words during course of word embedding generation.  
	Here, the pre-processed data is first converted to vectors with an ASCII supported integration. \par The dense vector representation of words learned by word2vec have shown remarkable progression in carrying semantic meanings\cite{meyer2016exactly}. Apparently, word2vec framework has been employed and vectors are neutrally trained separately for Radical and Non-Radical texts to generate word embeddings employing a Levenberg training mechanism. The validation criteria used while training: Propogative iterations and Gradient satisfaction. The weights are updated according to the equation:
	\begin{equation}
	    w_o=w_i * \alpha + b
	\end{equation}
	where $w_i$ is the input weight, $w_o$ is the output weight, learning rate $\alpha$ is set to default value of 1.0 and b is bias.
\begin{figure*} [!h]

	\includegraphics[width=\linewidth]{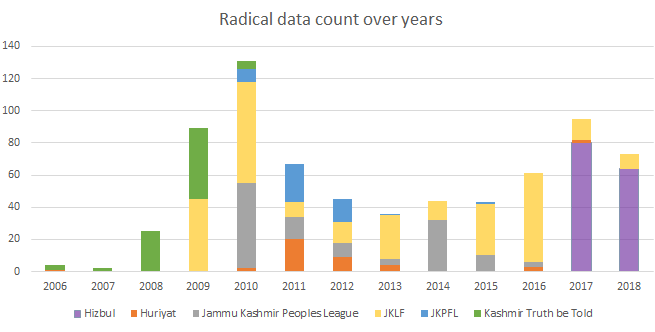}
	\caption{Radical texts timeline}
	\label{fig2}
\end{figure*}
	\item \textbf{LSTM and Fully-connected Layer}
	\\A feedforward neural network supported by LSTM is used for weight training as it is presumes texts to be interrelated.
	Many models use word embeddings as direct features while some use avg task i.e. calcuate the weighted average of word embeddings in similar context as representations of text\cite{lai2016generate}.
	Hence, the fully connected layer predicts labels for test datasets by comparing thresholds, generated using mean of radical and non-radical weights. 

	\item \textbf{Trend Analysis}
	\\The results of classification are visualized graphically which are discussed in detail in next section. 
\end{enumerate}

\section{Experimental Results}
We have choosen data from articles and blogs as they were identified radical by associated experts in the field of information warfare and therefore prodiguously contain radical content. There are total 1274 records in all after data cleaning where each record is comprised of 500 words on an average. The radical texts from the dataset are plotted over a timeline starting from 2006 till 2018 as shown in the Figure \ref{fig2}. It can be drawn that blog JKLF have been posting radical content for longer time than any other source, and Hizbul have the maximum radical data that have been posted in a year. 
\par
\begin{figure} [!h]

	\includegraphics[width=8cm]{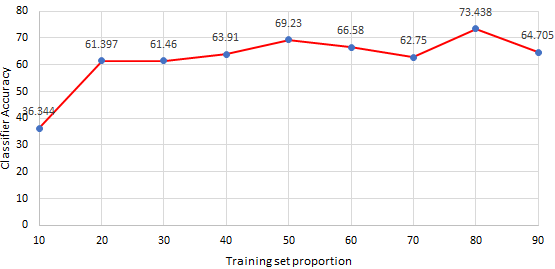}
	\caption{LSTM classifier accuracy at varied training set proportions.}
	\label{fig3}
\end{figure}

The results of the classification algorithm applied on different training set proportions can be seen in the Figure \ref{fig3} which implies maximum accuracy of 73.44\% is achieved with a 80:20 training and testing dataset ratio. The model starts to overfit beyond this ratio and accuracy drops as shown in Figure \ref{fig3}.
\par  As can be seen, the value of kappa coefficient achieved (here 0.796) indicates the level of agreement between two annotators in terms of data of documents collected. However, some authors of the blogs have mentioned their opinions from both perspectives: in favor of and against the radical ideology. Such texts lead to the computation of mean which may occur closer to any of the radical or nonradical means with slight changes in text. In such cases, the prediction model may go inaccurate and hence the accuracy drops.
\par
\begin{figure} [!h]

	\includegraphics[width=\linewidth]{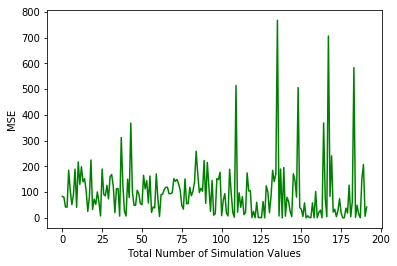}
	\caption{MSE vs Radical weights during the testing phase.}
	\label{fig4}
\end{figure}

 Further, the proposed approach is exercised over a 3-class classification for detection of irrelevant content in the data but accuracy declines to 34.23\% due to the presence of varied heterogeneous data in the irrelevant tagged section of data. 
The mean square error is plotted for radical content prediction while testing the model. The Figure \ref{fig4} exhibits the variations in prediction of radical texts on calculation of mean.
\par The proposed model is compared with few existing techniques for text analytics in order to establish the credibility of model in terms of different performance measures observed. At first, bag-of-words technique is employed to build a term document matrix. Then three machine learning algorithms namely SVM, Random Forest and MaxEnt are implemented to achieve classification of texts being radical or not radical. The algorithms applied have different ways of operating. Support Vector Machine focuses on finding a hyperplane in N-dimensional space that classifies the data points. Random forest takes into account many decision trees predicting classes individually, then the votes from all the trees are aggregated to decide final prediction of model. Maximum Entropy Classifier selects the model with largest entropy among all the models fitting training data for classification. The various performance measures being evaluated for comparison are enlisted in the Table \ref{measures}.
\begin{table}[!h]
\small
    \centering
    \begin{tabular}{|p{2cm}|p{1.5cm}|p{1.5cm}|p{1.5cm}|} 
     \hline
     \textbf{Approach} & \textbf{Precision} & \textbf{Recall} & \textbf{F-score} \\  [0.5ex] 
     \hline\hline
     
     Feedforward NN supported by LSTM & 85.96 & 53.26 & 65.77\\  [0.5ex] 
     \hline
     Random Forest & 73.50 & 64.00 & 65.55\\  [0.5ex] 
     \hline
     SVM & 53.50 & 50.50 & 45.00\\  [0.5ex] 
     \hline
     
     MaxEnt & 69.50 & 68.00 & 68.55
      \\[1ex] 
          \hline

    \end{tabular}
    \linebreak
    \caption{Different performance measures of our implementation expressed in percentage.}
    \label{measures}
\end{table}
\par
\begin{figure} [!h]

	\includegraphics[width=\linewidth]{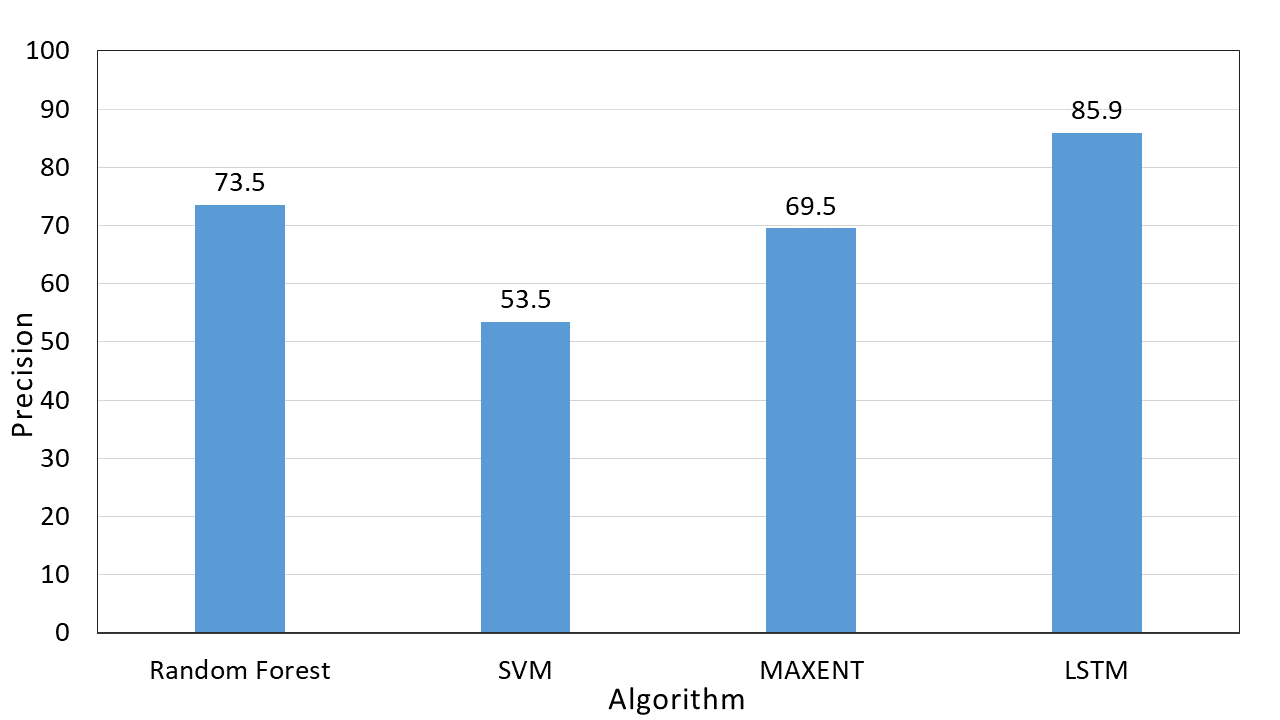}
	\caption{Comparison of precision score of different algorithms applied to the dataset.}
	\label{fig5}
\end{figure}

The comparison of precision score of different techniques is presented graphically in the Figure \ref{fig5} to demonstrate a contrasting view of implemented algorithms.

\section{Conclusion and Future Work}

The radical data detection task has been carried out in numerous studies in the past ranging from ontology-based techniques \cite{fernandez2018contextual} \cite{munezero2014automatic}, to machine learning algorithms. The feature extraction task before applying classification methods has utmost importance and requires plenty of effort. Even then it might miss out on capturing semantics of the text. Deep learning can be applied to overcome the drawbacks. In the present work, we have utilised LSTM based approach to identify radical content over online media. This model can help online social media sites in omission of inappropriate content. The proposed approach has been able to overstep most of the state-of-art techniques in terms of precision (85.9\%). This model can be extended with additional layer of CNN for explicit identification of features.

\section{Acknowledgements}
This work was partially supported by Cyber Security Research Centre,
Punjab Engineering College (Deemed to be University), Chandigarh, India.
The author Jaspal K Saini is grateful to Visvesvaraya PhD scheme for Electronics and IT
for funding this research.
\bibliographystyle{ACM-Reference-Format}
\bibliography{paper}

\end{document}